\documentclass[
    aps,
    fleqn,
    a4paper,
    superscriptaddress,
    preprint,
    preprintnumbers,
    showpacs,
    showkeys
]{revtex4}

\usepackage{graphicx}
\usepackage{tabularx}
\usepackage{supertabular}
\usepackage{picins}
\usepackage{amsmath}
\usepackage{amsfonts}
\usepackage[cp1250]{inputenc}

\newcommand{\De}[1][33]{\ensuremath{\Delta \varepsilon _{#1}}}
\newcommand{\Dd}[1][36]{\ensuremath{\Delta d _{#1}}}
\newcommand{\Ds}[1][66]{\ensuremath{\Delta s _{#1}}}

\newcommand{\KDP}{\ensuremath{{\rm KH_2PO_4}}}
\newcommand{\CDP}{\ensuremath{{\rm CsH_2PO_4}}}
\newcommand{\CDA}{\ensuremath{{\rm CsH_2AsO_4}}}
\newcommand{\RDP}{\ensuremath{{\rm RbH_2PO_4}}}
\newcommand{\BT}{\ensuremath{{\rm BaTiO_3}}}
\newcommand{\LTT}{\ensuremath{{\rm LiTlC_4H_4O_6\cdot H_2O}}}
\newcommand{\ALHS}{\ensuremath{{\rm (NH_4)_4LiH_3(SO_4)_4}}}

\newcommand{\Tc}{\ensuremath{T_{\rm C}}}

\newcommand{\1}[1]{#1^{(1)}}
\newcommand{\2}[1]{#1^{(2)}}

\newcommand{\dd}{{\rm d}}

\newcommand{\embox}[1]{{\em #1\/}}
\newcommand{\itbox}[1]{{\it #1\/}}
\newcommand{\bfbox}[1]{{\bf #1}}

\newcommand{\E}[1]{\!\cdot\! 10^{#1}}

\newcommand{\um}{{\rm \,\mu m}}
\newcommand{\m}{{\rm \,m}}
\newcommand{\C}{{\rm \,C}}
\newcommand{\Pa}{{\rm \,Pa}}
\newcommand{\J}{{\rm \,J}}

\newcommand{\tE}{\ensuremath{{\tau_{\rm ext}}}}
\newcommand{\eO}{\ensuremath{e_0}}
\newcommand{\PO}{\ensuremath{P_0}}
\newcommand{\sO}{\ensuremath{\sigma _0}}

\newcommand{\Uel}{\ensuremath{U_{\rm el}}}
\newcommand{\Udef}{\ensuremath{U_{\rm def}}}
\newcommand{\Uw}{\ensuremath{U_{\rm w}}}

\newcommand{\dWel}{\ensuremath{\delta W_{\rm el}}}
\newcommand{\dWdef}{\ensuremath{\delta W_{\rm def}}}

\begin{document}

\preprint{{\em Ferroelectrics\/}: Vol. 257, No. 1-4, Pgs. 211-220 (2001) - {\em (Revised version)}}

\title{
Extrinsic contributions in a nonuniform ferroic sample: Dielectric, piezoelectric and elastic
}

\author{P. \surname{Mokr\'{y}}}
\email{pavel.mokry@tul.cz}
\affiliation{Dept. of Physics, International Center for Piezoelectric Research, Technical University Liberec, Liberec 1, 461 17 Czech Republic}
\author{A. \surname{Kopal}}
\affiliation{Dept. of Physics, International Center for Piezoelectric Research, Technical University Liberec, Liberec 1, 461 17 Czech Republic}
\author{J. \surname{Fousek}}
\affiliation{Dept. of Physics, International Center for Piezoelectric Research, Technical University Liberec, Liberec 1, 461 17 Czech Republic}
\affiliation{Materials Research Laboratory, The Pennsylvania State University, State College, PA 16801, USA}

\date{8. September 2000}

\begin{abstract}
The contribution \De[]\ of extremely small motions of domain walls to
small-signal permittivity of a multidomain ferroelectric sample has been a
research issue for many years. In ferroelastic ferroelectrics such motions
contribute also to their piezoelectric (by \Dd[]) and elastic (by \Ds[])
properties. Data about their simultaneous existence are scarce but those
available point to mutual proportionality of \De[], \Dd[]\ and \Ds[], as
expected. To understand the magnitude of extrinsic contributions, the
origin of the restoring force acting on domain walls must be understood. In
the present contribution the theory has been developed based on the model of
a plate-like sample in which the ferroelectric- ferroelastic bulk is provided
with a nonferroic surface layer. Motion of domain walls in the bulk
results in a change of electric and elastic energy both in the bulk and in
the layer, which provides the source of restoring force. This makes it
possible to determine all mentioned extrinsic contributions. We discuss the
applicability of the model to available data for single crystals and also for
ceramic grains.
\end{abstract}


\maketitle

\section{Introduction}

The problem of extrinsic (domain wall) contributions has been
addressed by many authors, both experimentally and theoretically.
In the prevailing number of cases, only extrinsic permittivity has
been studied. For piezoelectric ceramics, Arlt et
al.\cite{ArtA3:Arlt1} were the first to address the problem of
wall contributions to all involved properties: permittivity
$\varepsilon$, elastic compliances $s$ and piezoelectric
coefficients $d$. In this and related papers, the existence of the
restoring force is assumed and its origin was not specified.
Later, Arlt and Pertsev\cite{ArtA3:Arlt2} offered a more
involved approach: when domain wall in a ceramic grain is shifted,
uncompensated bound charge appears on the grain boundary,
producing electric field. Simultaneously, if the involved domains
are ferroelastic, a wall shift results in mechanical stress in
surrounding grains. Thus the shift is accompanied by the increase
of both electric and elastic energies, leading to restoring
forces. Results of these theories have been successfully related
to experimental data on all three mentioned contributions: \De[],
\Ds[]\ and \Dd[].\cite{ArtA3:Arlt1,ArtA3:Arlt2}

While there exist a number of such data for ceramic
materials\cite{ArtA3:Arlt1,ArtA3:Zhang1,ArtA3:Zhang2},
information for single crystals is rather scarce. Understandably,
for nonferroelastic ferroelectrics such as TGS only data on \De[]\
are available;\cite{ArtA3:Fousek1} on the other hand for
crystals which are both ferroelectric and ferroelastic with more
than two domain states, dense domain systems are rather chaotical,
difficult to approach theoretically. In the present paper, we have
in mind ferroelectric and ferroelastic crystals with only two
domain states. In particular, crystals belonging to the \KDP\
family belong to this category and have been intensively studied.
Nakamura et al.\cite{ArtA3:Nakamura1,ArtA3:Nakamura2} determined
\De\ for \KDP, \CDP\ and \CDA. For \KDP, Nakamura and
Kuramoto\cite{ArtA3:Nakamura3} proved the existence of both \De\
and \Ds\ while \Dd\ was measured for \RDP\cite{ArtA3:Shuvalov1}.
For the same material, all quantities \De, \Dd\ and \Ds\ have been
measured by \v{S}tula et al.\cite{ArtA3:Stula1} It was found that
all these contributions are mutually proportional when measured as
a function of temperature, in the temperature interval between
\Tc\ and $\Tc$-35 K. Several other ferroics for which our approach
may be applicable will be mentioned at the end of this paper.

In single crystals, the origin of the restoring force is usually
connected to domain wall pinning on crystal lattice defects. In
our recent papers\cite{ArtA3:Kopal1,ArtA3:Kopal2} we introduced
the model of a passive surface layer to calculate the restoring
force for domain walls and the resulting extrinsic permittivity
and piezoelectric coefficient. In fact, the influence of a surface
layer on the properties of a ferroelectric sample was discussed
repeatedly several decades ago. In particular, in connection with
the sidewise motion of domain walls in \BT, thickness dependence
of the coercive field, asymmetry of a hysteresis loop or the
problem of energies of critical nuclei, theoretically impossibly
high.

In the present paper, we return to this approach. However, in contrast to
previous calculations, we offer a more involved model. The shift of a domain
wall induced by the application of electric field or elastic stress results
in the increase of \itbox{both} electric and elastic energies. In the
following, these are explicitly calculated which makes it possible to
determine all extrinsic coefficients \De, \Ds\ and \Dd. Their dependence on
the sample properties will be discussed.

\section{Description of the model}

%
\begin{figure}[t]
\begin{center}
\begin{minipage}{11cm}
  \includegraphics[width=85mm]{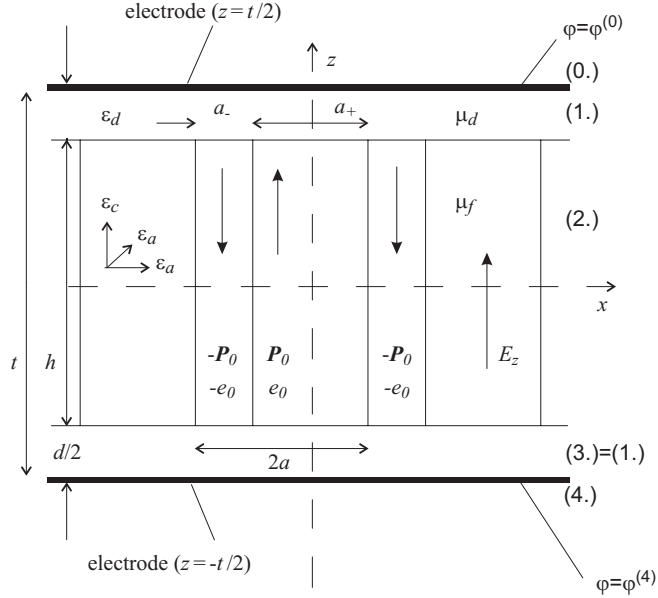}\vspace{3mm}
  \caption{Geometry of the model in $x$-$z$ plane
  \label{artA3:fig:xzgeometry}}
\end{minipage}
\end{center}
\end{figure}

%
\begin{figure}[t]
\begin{center}
\begin{minipage}{11cm}
  \includegraphics[width=85mm]{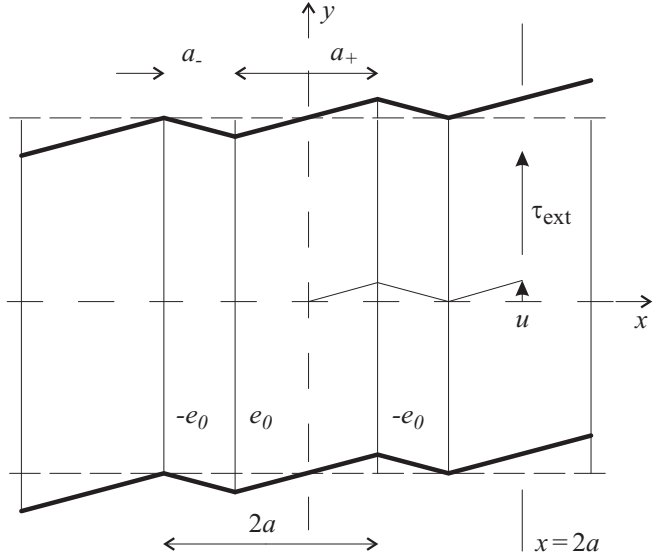}\vspace{3mm}
   \caption{Geometry of the model in $x$-$y$ plane \label{artA3:fig:xygeometry}}
\end{minipage}
\end{center}
\end{figure}
%
We consider a plate-like sample elecroded sample of infinite area
with major surfaces perpendicular to the ferroelectric axis $z$.
The material is both ferroelectric and feroelastic; domains with
antiparallel polarisation differ in the sign of spontaneous shear.
However, we shall approximate the material in the ferroelectric
phase by the equation of state
\begin{eqnarray}
    D_{x} &=& \varepsilon_0\varepsilon_a E_{x},
    \nonumber \\
    D_z &=& \varepsilon_0\varepsilon_c E_z + P_{0},
    \nonumber
\end{eqnarray}
neglecting the intrinsic piezoelectricity. As we shall see, this
does not supress the existence the extrinsic piezoelectricity,
which is one of the aims of our calculations. In the preceding
equation, we also neglect nonlinear terms; for 2nd order phase
transitions this limits the validity of our calculations to the
temperature region not very close bellow the temperature \Tc.
Domain walls are assumed to have surface energy density $\sigma_w$
and zero thickness.

For simplicity, we approximate elastically anisotropic material of
the sample by the elastically isotropic one. Neglecting again the
intrinsic piezoelectricity, its mechanical properties are
described by equations  for stress tensor components
\begin{eqnarray}
   \1{\tau_{ij}} &=& 2 \mu_d\1{e_{ij}}, \nonumber \\
   \2{\tau_{ij}} &=& 2 \mu_f\left(\2{e_{ij}} - e_{0,ij}\right), \nonumber
\end{eqnarray}
where $\mu_d$ and $\mu_f$ are Lame coefficients of the passive
layers and the bulk respectively; $e_{ij}$ is the strain tensor,
$e_{0,ij}$ is the spontaneous strain tensor of the central
ferroelastic part. We suppose that the only nonzero components of
the spontaneous strain tensor are $e_{0,12}=e_{0,21}=\pm\eO$ in
the $a_+$ resp. $a_-$ domain, see
Fig.\,\ref{artA3:fig:xygeometry}. We introduce the asymmetry
factor
\[
   A = \frac{a_+ - a_-}{a_+ + a_-}.
\]
We neglect thermal interactions and suppose that
the sample is thermally isolated. To keep the constant voltage $V$ on the
sample, the electrodes should be connected to external electrical source. In
the same way, to keep constant external stress $\tau_{{\rm ext,}12}=\tE$, the
sample should be deformed by external mechanical force. The infinitesimal
work of these external sources should be taken into consideration when
discussing the variations of the energy of the isolated system sample +
sources.

\section{Helmholtz free energy}

In what follows we consider three contributions to Helmholtz free
energy, calculated per unit area of the plate-like sample (in the
$x$-$y$ plane):

The energy of domain walls per unit area of the sample
\[
   \Uw=\sigma _w\, h/a \hspace{1cm}\left[{\rm J\cdot m^{-2}}\right],
\]
the electric field energy per unit area of the sample
\[
   \Uel(V,\, A)=\frac{1}2 \int_V E_{i}\, \left(D_{i}-P_{0,i}\right)\,
   \dd V,
\]
where the integration is taken over parallelepiped $x\in\left<0,\,
2a\right>$, $z\in\left<-t/2,\, t/2\right>$, $y\in\left<0,\,
1\,{\rm m^2}/2a\right>$, and energy of elastic deformations per
unit area of the sample:
\begin{eqnarray}
   \Udef(\tE,\, A)&=&\frac{1}2 \int_V \tau_{ij}\,
   \left(e_{ij}-e_{0,ij}\right)\,
   \dd V, \mbox{\qquad resp.} \nonumber \\
   \Udef(u,\, A)&=&\frac{1}2 \int_V \tau_{ij}\, \left(e_{ij}-e_{0,ij}\right)\,
   \dd V,  \nonumber
\end{eqnarray}
where the integration is taken over the same region, in the first
case for constant external stress \tE\ in the plane $x=2a$, in the
second one for constant displacement $u$ in the plane $x=2a$. In
both cases, the displacement for $x=0$ is chosen to be zero and
the boundaries of the sample in x-y plane are free of stress.

To find the \Uel\ and \Udef, we have calculated electric potential
and mechanical displacement inside the sample by Fourier method.
We present here only the relative simple results for a ``dense
pattern approximation", that is for $d,\, h \gg a$:
\begin{eqnarray}
    \Uel(V,\, A) &=&
        \frac{
            dh\, A^2 \PO^2 + \, V^2\,\varepsilon_0 ^2 \varepsilon _d\varepsilon _c
        }{
            2 \varepsilon _0
            \left(\varepsilon_d\,h + \varepsilon _c\,d\right)
        },
    \nonumber \\
    \Udef(\tE,\, A) &=&
        \frac{
            ht^2 \mu_f \tau_{\rm ext}^2
        }{
            2\left(\mu_d\, d + \mu_f\, h\right)^{2}
        }
        +
        \frac{d \mu_d}{2}
        \left(
            2A\eO + \frac{t \tE}{\mu_d\, d + \mu_f\, h}
        \right)^2,
    \nonumber \\
    \Udef(u,\, A) &=&
        \frac{
            \mu_d du^2 + \mu_f h\left(2 u - 8A\eO a\right)^2
        }{
            8a^2
        }. \nonumber
\end{eqnarray}

Infinitesimal work of the electric source at constant voltage $V$ per unit
area of the sample plate is
\[
   \dWel = V\, \delta \sO \hspace{1cm}\left[{\rm J\cdot m^{-2}}\right],
\]
where \sO\ is constant Fourier component of the free charge density on the
positive electrode, calculated as
\begin{equation}
    \sO(V,\, A)=
        \frac{
            \PO A + \varepsilon _0 \varepsilon_c\, V/h
        }{
            1 + \varepsilon_c d/\left(\varepsilon_d h\right)
        }.
   \label{artA3:eq:sigma0}
\end{equation}
Infinitesimal work of the mechanical source, deforming the parallelepiped is
\[
   \dWdef =\tE\, t/(2a)\, \delta u
   \hspace{1cm}\left[{\rm J\cdot m^{-2}}\right],
\]

It is easy to prove that the equilibrium domain structure for
$V=0$, $\tE=0$ resp. $u(x=2a)=0$ is symmetric (i.e. A=0), with
domain width
\begin{equation}
    a_{\rm eq} =
        \sqrt{3.68\ h\sigma_w}
        \left[
            4 e_0^2\,
            \frac{\mu_d\mu_f}{\mu_d+\mu_f}
            +
            \frac{
                P_0^2
            }{
                \varepsilon _0
                \left(
                    \varepsilon_d + \sqrt{\varepsilon_a\, \varepsilon_c}
                \right)
            }
        \right]^{-\frac{1}2}.
  \label{artA3:eq:weq}
\end{equation}

It can be shown (see e.g.\cite{ArtA3:Kopal1}) that within a
large interval of the applied voltage the average width
\[
   a = \left(a_+ + a_-\right)/2
\]
remains constant. This is why \Uw\ can be also considered as constant.

\section{Extrinsic contributions \De, \Ds\ and \Dd}

We calculate the equilibrium effective $\varepsilon ^{\rm eff} _{33}$ of the
sample from the relations
\begin{equation}
   D ^{\rm eff}_3 = \sO(V,\, A) = \varepsilon ^{\rm eff} _{33}\, E ^{\rm ext} _3
   = \varepsilon ^{\rm eff} _{33}\, \frac{V}{t}.
   \label{artA3:eq:Deff3rel}
\end{equation}
We keep $\tE=0$, $V$=constant and we take into account that
variation of ``Helmholtz free energy of the sample + the work of
electric source" is zero in equilibrium:
\[
   \frac{\partial \Uel(V,\, A)}{\partial A}\, \delta A +
   \frac{\partial \Udef(\tE=0,\,A)}{\partial A}\, \delta A -
   \dWel = 0.
\]
Solving this standard problem, we get $A=A(V)$ and from the Eqs.
(\ref{artA3:eq:sigma0}), (\ref{artA3:eq:Deff3rel}) the effective
$\varepsilon ^{\rm eff} _{33}$. For $\mu_d=\mu_f=\mu$,
$\varepsilon _d=\varepsilon _c=\varepsilon _z$ we obtain the
relatively simple result:
\begin{equation}
    \varepsilon ^{\rm eff} _z =
        \varepsilon _z
        +
        \varepsilon _z\,
        \frac{h}{d}\, \cdot\,
        \left[
            \frac{
                P_0^2\, h
            }{
                P_0^2\, h + 4\, e_0^2\, \varepsilon _0\, \varepsilon _z\, \mu t
            }
        \right] .
   \label{artA3:eq:epseff}
\end{equation}

The effective elastic compliance of the sample is
\[
   s^{\rm eff} _{66} = 4 s^{\rm eff} _{1212} = \frac{2 e^{\rm eff}
   _{12}}{\tE} = \frac{u}{2\, a\, \tE}.
\]
We put $V=0$ (shorted sample), apply constant external shear
stress \tE\ and postulate, that variation of ``Helmholtz free
energy of the sample + the work of mechanical source" is zero in
equilibrium
\[
   \frac{\partial \Uel(0,\, A)}{\partial A}\, \delta A +
   \frac{\partial \Udef(u,\, A)}{\partial A}\, \delta A -
   \dWdef = 0.
\]
Solving this problem, we get $u=u(\tE)$ and we get for
$\mu_d=\mu_f=\mu$, $\varepsilon _d=\varepsilon _c=\varepsilon _z$:
\begin{equation}
    s^{\rm eff} _{66} =
        \frac{1}{\mu}
        +
        \frac{1}{\mu}\,\cdot
        \frac{h}{d} \,
        \left[
            \frac{
                4\, e^2_0\, \varepsilon _0\, \varepsilon _z
            }{
                (P^2_0/\mu)
                +
                4\, e^2_0\, \varepsilon _0\, \varepsilon _z
            }
        \right].
   \label{artA3:eq:seff}
\end{equation}

To find the effective piezoelectric coefficient of the sample
\[
   d^{\rm eff} _{36} = \frac{ D^{\rm eff} _{3}}{\tau _{\rm ext,6}} =
   \frac{\sO}{\tau _{\rm ext,6}},
\]
we put $V=0$, apply \tE\ and solve the problem
\[
   \frac{\partial \Uel(0,\, A)}{\partial A}\, \delta A +
   \frac{\partial \Udef(u,\, A)}{\partial A}\, \delta A +
   \frac{\partial \Udef(u,\, A)}{\partial u}\, \delta u -
   \dWdef = 0.
\]
From here we obtain $A=A(\tE)$. Inserting this result into Eq.
(\ref{artA3:eq:sigma0}) we obtain $\sO=\sO(\tE)$. For
$\mu_d=\mu_f=\mu$, $\varepsilon _d=\varepsilon _c=\varepsilon _z$
we get finally for the effective piezoelectric coefficient
\begin{equation}
    d^{\rm eff} _{36} =
        \frac hd \,\cdot\,
        \left[
            \frac{
                2 \eO\, \PO\, \varepsilon _0\, \varepsilon _z
            }{
                P^2_0 + 4\, e^2_0\, \varepsilon _0\, \varepsilon _z\, \mu
            }
        \right].
   \label{artA3:eq:deff}
\end{equation}
The same result we get for the inverse piezoelectric effect.

\section{Discussion}

\begin{table}[t]
\renewcommand{\arraystretch}{1.3}
\tabcolsep=8mm \caption{ {Numerical estimate of \De,\, \Ds\
and \Dd\ for different values of surface layer thickness. The
following numerical constants have been used: $\PO=4
\E{-2}\C\m^{-2}$, $\varepsilon _z=100$, $\eO=0.015$,
$\mu=6\E{9}\Pa$, $\sigma _w=5\E{-3}\J\m^{-2}$, $h=5\E{-4}\m$.
\label{artA3:tab:NumResults} } }
\begin{supertabular*}{\textwidth}{l|l|l|l}
      \hline \hline
         \hfill &$d=0.5\E{-6}\m$ &$d=2.5\E{-6}\m$& $d=5\E{-6}\m$ \\
      \hline
         $\De\, \left[1\right]$& 12\,000& 2\,500& 1\,200 \\
         $\Dd\, \left[\C\m^{-2}\right]$& $4.1\E{-8}$& $8.3\E{-9}$& $4.1\E{-9}$ \\
         $\Ds\, \left[\Pa^{-1}\right]$ &$1.6\E{-8}$ &$3.2\E{-9}$ &$1.6\E{-9}$
         \\
      \hline \hline
\end{supertabular*}
\end{table}
The above calculation leads to explicite formulae
(\ref{artA3:eq:epseff}) to (\ref{artA3:eq:deff}) for extrinsic
permittivity, extrinsic elastic compliance and extrinsic
piezoelectric coefficient, resp. Numerical values for all involved
material coefficients are available for single crystals of \RDP:
$\PO=4\E{-2}\C\m^{-2}$, $\varepsilon _z=100$, $\eO=0.015$ and
$\mu=6\E{9}\Pa$. To obtain numerical estimates for particular
samples we put $\sigma _w=5\E{-3}\J\m^{-2}$ and $h=5\E{-4}\m$ and
choose three values of surface layer thickness, namely
$d=0.5\E{-6}\m$, $d=2.5\E{-6}\m$ and $d=5\E{-6}\m$. Table
\ref{artA3:tab:NumResults} shows resulting values of all three
extrinsic variables. These numbers appear very realistic and
confirm the applicability of the model presented in this paper.

It is appropriate to pay some attention to the fact that also the
formula (\ref{artA3:eq:weq}) gives a reasonable numerical
estimation for the width of equilibrium domain pattern. With
numerical values specified at Tab. \ref{artA3:tab:NumResults} we
obtain $a_{\rm eq} \cong 1\um$.

In the approach analyzed above, the source of the restoring force
acting on domain walls is the interaction of ferroic sample with a
passive surface layer. Very often, the origin of restoring forces
is connected with domain wall pinning to defects. Understandably,
the latter mechanism cannot be excluded for ferroics of any
chemical composition. On the other hand, passive surface layers
can be formed during sample preparations and in particular for
water-soluble crystals their appearance is a very likely: samples
are polished in water-containing media, the procedure obviously
leading to the presence of a passive surface layer. The example
analyzed numerically above, crystals of \RDP, falls into this
category. However, extrinsic properties of a number of other
crystals have been studied. Thus, e.g., for \LTT\ (species
222-$P\varepsilon ds$-$2_y$) \cite{ArtA3:Sawaguchi1} very strong
and nonhysteretic dependence of $s_{44}^{\rm E}$ as a function of
applied field $E$ was measured. This is possible to explain by a
strong contribution of domain walls with a pronounced restoring
force. The above model would lead to such behavior. Similarly,
large extrinsic contributions to $s_{11}$ have been
measured\cite{ArtA3:Zimmermann1} for \ALHS, species
4-$\varepsilon ds$-2.

\acknowledgements

This work has been supported by the Ministry of
Education of the Czech Republic (Projects No.~VS~96006 and
No.~MSM~242200002) and by the Grant Agency of the Czech Republic
(Project No.~202/00/1245).

\end{document}